\documentclass[aps,prb,twocolumn,superscriptaddress,showpacs,preprintnumbers,amsmath,amssymb]{revtex4}
\usepackage{amsmath}
\usepackage{dsfont}
\usepackage{amstext}

\usepackage{amssymb}
\usepackage{amsbsy}
\usepackage{amsthm}
\usepackage{epsfig}
\usepackage{graphicx}
\usepackage{bbm}
\usepackage{hyperref}
\usepackage{color}

\newcommand{\ie}{{\emph{i.e.~}}}
\makeatletter

\newcommand{\Rmnum}[1]{\expandafter\@slowromancap\romannumeral #1@}
\makeatother
\newcommand{\imth}{\hspace{1pt}\mathrm{i}\hspace{1pt}}

\newcommand{\eg}{{\emph{e.g.~}}}

\newcommand{\mbz}{{\mathbb{Z}}}
\newcommand{\Tr}[1]{\text{Tr}\Big[{#1}\Big]}
\newcommand{\bpm}{\begin{pmatrix}}
\newcommand{\epm}{\end{pmatrix}}
\newcommand{\bea}{\begin{eqnarray}}
\newcommand{\eea}{\end{eqnarray}}

\def\triangpic{{\begin{picture}(12,14)(-1,-1)
                      \put (0,0) {\circle*{3}}
		      \put (0,12) {\circle*{3}}
		      \put (10,6) {\circle*{3}}
		      \put (0,0) {\line (0,1) {12}}
		      \put (0,12) {\line (5,-3) {10}}
		      \put (0,0) {\line (5,3) {10}}
                \end{picture}}}

\def\nnring{{\begin{picture}(22,8)(-1,-7)
                      \put (0,0) {\circle*{3}}
		      \put (20,0) {\circle*{3}}
		      \put (10,-6) {\circle*{3}}
		      \put (0,0) {\vector (1,0) {20}}
		      \put (20,0) {\line (-5,-3) {10}}
		      \put (0,0) {\line (5,-3) {10}}
                \end{picture}}}

\def\nnnring{{\begin{picture}(22,19)(-1,-1)
                      \put (0,0) {\circle*{3}}
		      \put (20,0) {\circle*{3}}
		      \put (10,15) {\circle*{3}}
		      \put (0,0) {\vector (1,0) {20}}
		      \put (20,0) {\vector (-2,3) {10}}
		      \put (10,15) {\vector (-2,-3) {10}}
                \end{picture}}}

\begin{document}
\title{Spin quantum Hall effects in a spin-1 topological paramagnet}

\author{Yuan-Ming Lu}
\affiliation{Department of Physics, University of California, Berkeley, CA 94720, USA}
\affiliation{Materials Science Division, Lawrence Berkeley National Laboratories, Berkeley, CA 94720}

\author{Dung-Hai Lee}
\affiliation{Department of Physics, University of California, Berkeley, CA 94720, USA}
\affiliation{Materials Science Division, Lawrence Berkeley National Laboratories, Berkeley, CA 94720}

\begin{abstract}
AKLT state (or Haldane phase) in a spin-1 chain represents a large class of gapped topological paramagnets, which hosts symmetry-protected gapless excitations on the boundary. In this work we show how to realize this type of featureless spin-1 states on a generic two-dimensional lattice. These states have a gapped spectrum in the bulk but supports gapless edge states protected by spin rotational symmetry along a certain direction, and are featured by spin quantum Hall effect. Using fermion representation of integer-spins we show a concrete example of such spin-1 topological paramagnets on kagome lattice, and suggest a microscopic spin-1 Hamiltonian which may realize it.
\end{abstract}

\maketitle


\section{Introduction}

The fate of frustrated quantum magnets at very low temperatures has for long been an important question in condensed matter physics\cite{Diep2005B,Lacroix2011B}. In contrast to unfrustrated magnets where all spins freeze into an ordered pattern below a certain critical temperature, geometric frustrations and/or quantum fluctuations could strongly suppress the long-range order in frustrated magnets. When these frustrations/fluctuations dominate, magnets refuse to order even down to zero temperature and form a liquid-like cooperative paramagnet. One possibility for such a zero-temperature disordered many-spin state is the quantum spin liquid\cite{Lee2008a,Balents2010}, which supports bulk quasiparticles obeying fractional statistics\cite{Wilczek1990B,Kalmeyer1987,Wen2002}. Another possibility is a non-fractionalized featureless spin state, which is gapped in the bulk but hosts exotic gapless excitations on the boundary. A famous example of these ``topological paramagnets'' is the AKLT state\cite{Affleck1987,Affleck1987a} in a spin-1 chain\cite{Haldane1983a}. On each end of such an open chain there is a spin-half excitation. Recently it was revealed that the stability of such topological paramagnets (and their boundary excitations) are protected by certain symmetry\cite{Gu2009,Pollmann2012}. In other words, in the absence of any symmetry, such a AKLT-type state can be continuously connected to a trivial product state (with no boundary excitations) without closing the bulk energy gap. These non-fractionalized gapped phases are hence coined\cite{Chen2013} ``symmetry protected topological phases''. For instance AKLT state in one spatial dimension can be protected by either $SO(3)$ spin rotational symmetry or time reversal symmetry\cite{Pollmann2012}.

IA natural question is: in two spatial dimensions, do similar topological paramagnets exist for integer spins, analogous to one-dimensional AKLT state (or Haldane phase)? The answer is yes. In fact there are an infinite number of different gapped non-fractionalized integer-spin phases, which have gapless edge excitations protected by either $SO(3)$ or $U(1)$ spin rotational symmetry\cite{Chen2013,Liu2013,Senthil2013,Ye2013,Oon2013}. They are also featured by spin quantum Hall effect\cite{Senthil1999,Liu2013}, \ie quantized Hall conductance $\sigma_{xy}^s=2k,~k\in\mbz$ in unit of $\hbar/2\pi$ for spins. To be specific, the spatial gradient of external Zeeman field $B_z$ in one direction would result in spin current $\vec{j}^{S^z}$ in the perpendicular direction:
\begin{eqnarray}
j^{S^z}_x=-\sigma_{xy}^s\frac{\text{d}B_z(y)}{\text{d}y}.
\end{eqnarray}
The gapless edge states and quantized response $\sigma_{xy}^s$ will be protected as long as $U(1)_{S^z}$ (spin rotation along $\hat{z}$-axis) symmetry is preserved. Notice that \emph{spin quantum Hall effect} here is different from the quantum spin Hall effect\cite{Kane2005,Bernevig2006,Konig2007} in time reversal invariant systems, which on the other hand measures the spin response to a transverse electric field. Similar gapped featureless boson states protected by $U(1)$ symmetry (boson charge conservation) have been proposed in the continuum\cite{Lu2012a,Senthil2013} and on the lattice\cite{Lu2014f,Grover2013}. However so far a lattice realization of these featureless non-fractionalized states in a integer-spin system is still missing.

In this work we propose a general way to realize these AKLT-like topological paramagnets protected by $U(1)_{S^z}$ symmetry on a two-dimensional lattice. We use the fermion representation\cite{Liu2010a} of integer spins to construct their many-body wavefunctions. We also derive low-energy effective theory of these featureless gapped phases and their spin quantum Hall responses with $\sigma_{xy}^s=\pm2$. A careful analysis shows that their edge excitations will remain gapless, robust against any perturbations, as long as $U(1)_{S^z}$ symmetry is not broken. A concrete example of spin-1 magnets on kagome lattice is provided to demonstrate the general construction: the corresponding spin-1 topological paramagnet preserves all kagome lattice symmetries as well as $U(1)_{S^z}$ spin rotational symmetry. Such a state has no net magnetization. Moreover a microscopic Hamiltonians which may realize these states on kagome lattice is proposed.


\section{Fermionic representation of $S=1$ spins}

For arbitrary spin-$S$ one can introduce $2S+1$ species of fermions $\{f_m|-S\leq m\leq S\}$ to represent the spin operator\cite{Liu2010,Liu2010a}:
\begin{eqnarray}\label{fermion rep:S}
\hat{\bf S}=f^\dagger{\bf I}f,~~~f\equiv(f_{S},f_{S-1},\cdots,f_{-S})^T.
\end{eqnarray}
where three $(2S+1)\times(2S+1)$ matrices $I^a$ are given by
\begin{eqnarray}
I^a_{m,n}=\langle S,m|\hat{S}^a|S,n\rangle,~~~a=x,y,z.\notag
\end{eqnarray}
$|S,m\rangle$ stands for the $S^z=m$ eigenstate of a spin-$S$. In the case of spin-1 \eg we have
\begin{eqnarray}
\notag&\hat{S}^+\equiv\hat{S}^x+\imth\hat{S}^y=\sqrt2(f^\dagger_{+1}f_0+f^\dagger_{0}f_{-1}),\\
&\hat{S}^z=f^\dagger_{+1}f_{+1}-f^\dagger_{-1}f_{-1}.\label{fermion rep:S=1}
\end{eqnarray}
The Hilbert space of these $2S+1$ species of fermions ($2^{2S+1}$-dimensional) is larger than the physical $(2S+1)$-dimensional Hilbert space of a spin-$S$, therefore we need to project the fermion system into the physical Hilbert space of spins. Such a projection is enforced by the following single-occupancy constraint:
\begin{eqnarray}\label{constraint:S=1}
\hat{N}_f\equiv\sum_{m=-S}^Sf^\dagger_mf_m=1.
\end{eqnarray}
In other words the physical spin-$S$ state $|spin\rangle$ is obtained by projection ${\bf P}(\hat{N}_f=1)\equiv\prod_{\bf r}\Big(1-\sum_mf^\dagger_{{\bf r},m}f_{{\bf r},m}\Big)$
\begin{eqnarray}
|spin\rangle={\bf P}(\hat{N}_f=1)|fermion\rangle.\label{projection:S}
\end{eqnarray}
on fermion ``mean-field'' state $|fermion\rangle$. More specifically, the many-body wavefunction $\Phi\big(S^z_{\bf r}=m_{\bf r}\big)$ for spin-$S$ operators $\hat{\bf S}_{\bf r}=\sum_{m,n}f^\dagger_{{\bf r},m}{\bf I}_{m,n}f_{{\bf r},n}$ on a lattice ({\bf r} denotes lattice sites) is given by
\begin{eqnarray}
\Phi\big(\{S^z_{\bf r}=m_{\bf r}\}\big)=\langle0|\prod_{\bf r}f_{{\bf r},m_{\bf r}}|fermion\rangle.\label{wf:S}
\end{eqnarray}
where $|0\rangle$ denotes the vacuum for fermions $\{f_m\}$. The fermion state $|fermion\rangle$ must have the proper filling: \ie one fermion per site on average, otherwise the projection (\ref{projection:S}) or (\ref{wf:S}) will vanish identically.





\section{Many-body wavefunction and effective theory}

Under a spin rotation $\hat{R}^z_\theta\equiv e^{\imth\theta\hat{S}^z}$ along $\hat{z}$-axis by an angle $\theta$, the fermions $\{f_m\}$ transform as
\begin{eqnarray}\label{symmetry:U(1)}
f_m\rightarrow(\hat{R}^z_\theta)^\dagger f_m \hat{R}^z_\theta=e^{\imth m\theta}f_m.
\end{eqnarray}
up to a $U(1)$ gauge redundancy\footnote{Notice that all spin operators in (\ref{fermion rep:S}) are invariant under this $U(1)$ gauge transformation: $f_m\rightarrow e^{\imth\phi}f_m,~~~\phi\in[0,2\pi)$ for $\forall~m$.} $f_m\rightarrow e^{\imth\phi}f_m,~~~\phi\in[0,2\pi)$.

We construct the fermion state $|fermion\rangle$ by filling energy levels of the following hopping Hamiltonian of fermions: $H_{f}=\sum_{m,n=-S}^{+S}\sum_{{\bf r},{\bf r}^\prime}f^\dagger_{{\bf r},m}\mathcal{H}^{m,n}_{{\bf r},{\bf r}^\prime}f_{{\bf r}^\prime,n}$.
Notice that the number of fermions in the mean-field ground state $|fermion\rangle$, \ie the number of filled single-particle energy levels, is equal to the total number of lattice sites, in order to guarantee the constraint (\ref{constraint:S=1}) on every site. In the presence of $U(1)_{S^z}$ spin rotational symmetry along $\hat{z}$-axis, those hopping terms mixing different species of fermions are forbidden by symmetry (\ref{symmetry:U(1)}) and the above hopping Hamiltonian reduces to $\mathcal{H}^{m,n}_{{\bf r},{\bf r}^\prime}=\delta_{m,n}\mathcal{H}^m_{{\bf r},{\bf r}^\prime}$
\begin{eqnarray}
H_{f}=\sum_{m=-S}^{+S}\sum_{{\bf r},{\bf r}^\prime}f^\dagger_{{\bf r},m}\mathcal{H}^{m}_{{\bf r},{\bf r}^\prime}f_{{\bf r}^\prime,m}\label{mean-field Hamiltonian}
\end{eqnarray}
In order to obtain a gapped spin-$S$ state, we require that each species of fermions $f_m$ to form a band insulator respectively. Constraint (\ref{constraint:S=1}) implies that on average the filling number of each species of fermions $f_m$ is $\nu=1/(2S+1)$ per site.

Now let's focus on a spin-1 system on a lattice $\{{\bf r}\}$, which involves 3 species of fermions $\{f_{{\bf r},+1};f_{{\bf r},0};f_{{\bf r},-1}\}$ in the fermion representation (\ref{fermion rep:S=1}). Since each species of fermions $f_{{\bf r},m}$ has a filling $\nu=1/3$ per site and forms a band insulator, there is a well-defined Chern number\cite{Thouless1982} $C_m$ (or Hall conductance) for each species of fermions. In the following we derive the effective theory of the spin-$S$ state obtained from projection (\ref{wf:S}) on the fermion state. For simplicity we first assume the total Chern number for all filled bands of fermion species $f_m$ is $C_m=\pm1$. Conserved fermion currents $J^\mu_m$ can be expressed in terms of dynamical $U(1)$ gauge fields $a_\mu^m$ as
\bea
J^\mu_m=\frac{\epsilon^{\mu\nu\lambda}}{2\pi}\partial_\nu a_\lambda^m,~~~m=-S,-S+1,\cdots,S-1,S.\notag
\eea
by non-relativistic duality transformation\cite{Fisher1989,Zee2010B} in $2+1$-dimensions. Summation over repeated indices $\mu,\nu,\lambda=t,x,y$ is always assumed. The fermion band structure is described by the following effective Chern-Simons theory:
\begin{eqnarray}
\notag\mathcal{L}_{f}=-\frac{\epsilon^{\mu\nu\lambda}}{4\pi}\sum_{m=-S}^SC_m\cdot a^m_\mu\partial_\nu a^m_\lambda,~~~C_m=\pm1.
\end{eqnarray}
The local constraint (\ref{constraint:S=1}) can be written in a covariant form:
\begin{eqnarray}
\notag\frac{\epsilon^{\mu\nu\lambda}}{2\pi}\sum_m\partial_\nu{a}^m_\lambda=\sum_{m=-S}^SJ_m^\mu=\bar{J}^\mu\equiv\frac{\epsilon^{\mu\nu\lambda}}{2\pi}\partial_\nu\bar{a}_\lambda
\end{eqnarray}
where $\bar{a}_\mu$ is a non-dynamical (constant) background field, whose field strength $\bar{J}^0=(\partial_x\bar{a}_y-\partial_y\bar{a}_x)/2\pi$ is a constant \Big(r.h.s. of Eq. (\ref{constraint:S=1})\Big). Note that due to constraint (\ref{constraint:S=1}) the fermions $f_m$ are not free fermions: they couple to a dynamical $U(1)$ gauge field. The constraint (\ref{constraint:S=1}) can be implemented by introducing $U(1)$ gauge field $b_\mu$ serving as a Lagrangian multiplier:
\begin{eqnarray}
\mathcal{L}_{constraint}=\frac{\epsilon^{\mu\nu\lambda}}{2\pi}b_\mu\partial_\nu\Big(\sum_{m=-S}^S a_\lambda^m-\bar{a}_\lambda\Big).\notag
\end{eqnarray}
After integrating out the gauge fields $b^\mu$ and $a^0_\mu$,~
we can obtain the low-energy long-wavelength effective theory of an $S=1$ (spin-$1$) state (\ref{wf:S}):
\begin{eqnarray}
&\mathcal{L}_{CS}=\mathcal{L}_{f}+\mathcal{L}_{constraint}=-\frac{\epsilon^{\mu\nu\lambda}}{4\pi}\begin{pmatrix}a^{+1}_\mu\\ a^{-1}_\mu\end{pmatrix}^T{\bf K}~\partial_\nu\begin{pmatrix}a^{+1}_\lambda\\ a^{-1}_\lambda\end{pmatrix}\notag
\end{eqnarray}
where $2\times2$ matrix ${\bf K}$ is given by
\begin{eqnarray}\label{K mat:S=1}
{\bf K}=\begin{pmatrix}C_{+1}+C_0&C_0\\C_0&C_{-1}+C_0\end{pmatrix}
\end{eqnarray}
In the presence of $U(1)_{S^z}$ spin rotational symmetry, the spin current of $S^z$ component is conserved and we can couple it to an external $U(1)$ ``spin gauge field'' $A^s_\mu$:
\begin{eqnarray}\label{lagrangian:S=1}
\mathcal{L}_{eff}=\mathcal{L}_{CS}-\frac{\epsilon^{\mu\nu\lambda}}{2\pi}A_\mu^s\partial_\nu\Big(\sum_{m=-S}^Sm\cdot a^m_\lambda\Big).
\end{eqnarray}
By integrating out the dynamical gauge fields $a^{\pm1}_\mu$, one can obtains the quantized Hall response for spin ($S^z$) transport of the $S=1$ state:
\begin{eqnarray}
&\mathcal{L}_{response}=\frac{C_{+1}+C_{-1}+4C_0}{C_{+1}C_{-1}+C_0(C_{+1}+C_{-1})}\cdot\frac{\epsilon^{\mu\nu\lambda}}{4\pi}A_\mu^s\partial_\nu A^s_\lambda\label{spin quantum hall}
\end{eqnarray}\\
In other words the spin quantum Hall conductance is $\sigma^{s}_{xy}=\frac{C_{+1}+C_{-1}+4C_0}{C_{+1}C_{-1}+C_0(C_{+1}+C_{-1})}$ in unit of $\frac1{2\pi}$ for this spin-1 state, assuming the Chern numbers take value of $C_m=\pm1$. As will be discussed later, gapless edge states protected by $U(1)_{S^z}$ symmetry are responsible for the spin quantum Hall effect\cite{Liu2013} here.

Notice that for a spin-1 state described by effective theory (\ref{lagrangian:S=1}), its ground state degeneracy on a torus\cite{Wen1990a,Keski-Vakkuri1993} is $|\det{\bf K}|$ in (\ref{K mat:S=1}). For the case $C_{+1}=C_{-1}=-C_0=\pm1$, we have $\det{\bf K}=-1$ and hence all the excitations are bosons (\ie no fractionalization) in the corresponding spin-1 state (\ref{wf:S}). Therefore it is a gapped featureless non-fractionalized spin-1 state with $U(1)_{S^z}$ symmetry. Besides it has gapless edge excitations protected by $U(1)_{S^z}$ symmetry\cite{Lu2012a,Senthil2013,Chen2012,Liu2013}, responsible for the spin quantum Hall conductance $\sigma_{xy}^s=\pm2$.

It's also worth mentioning that when $\mathcal{H}^m_{{\bf r},{\bf r}^\prime}\equiv\mathcal{H}^0_{{\bf r},{\bf r}^\prime},~\forall m$ in fermion hopping Hamiltonian (\ref{mean-field Hamiltonian}), the projected spin-$S$ state (\ref{wf:S}) has full $SO(3)$ spin rotational symmetry. When every fermion species has the same Chern number $C_{+1}=C_{-1}=C_0=\pm1$, the resultant spin-1 state (\ref{wf:S}) is a chiral spin liquid\cite{Kalmeyer1987} with
\bea
{\bf K}=\pm\bpm2&1\\1&2\epm.
\eea
in its effective theory (\ref{lagrangian:S=1}). It has two chiral edge modes and spin quantum Hall effect ($\sigma_{xy}^s=\pm2$), as well as fractional excitations in the bulk. To be specific there are two different anyon excitations: both have statistical angle $\theta=\frac{2\pi}{3}$, but their mutual braiding statistics is $\theta^\prime=\frac{2\pi}3$.

There is a simple physical picture which explains why the choice of Chern numbers $C_{+1}=C_{-1}=-C_0=\pm1$ gives rise to a non-fractionalized topological paramagnet. A spin-1 system can always be mapped to a boson system consisting of two species of hard-core bosons $\gamma_+$ and $\gamma_-$: they carry spin $S^z=\pm1$ respectively. A $S^z=0$ spin is mapped to a lattice site with no boson occupancy. The choice of Chern number $C_0=\pm1$ simply determines the flux attached to each boson to form composite fermions\cite{Jain1989}. These composite fermions are nothing but $f_{+1}$ and $f_{-1}$ in representation (\ref{fermion rep:S=1}). A natural question is: imagine we insert an external $2\pi$ flux $\Phi_0$ coupled to boson $\gamma_+$ (and hence to composite fermion $f_{+1}$), what's the response of the system? It's straightforward to work out that external flux $\Phi_0$ induces local charge polarizations
\bea
&\notag\delta\rho_{+1}/\Phi_0=\frac{C_{+1}+C_0}{C_{+1}C_{-1}+C_0(C_{+1}+C_{-1})},\\
&\notag\delta\rho_{-1}/\Phi_0=\frac{-C_0}{C_{+1}C_{-1}+C_0(C_{+1}+C_{-1})}.
\eea
Notice that a fractional charge polarization will imply nontrivial ground state degeneracy\cite{Oshikawa2006}, hence we need to choose Chern numbers satisfying
\bea
C_{+1}C_{-1}+C_0(C_{+1}+C_{-1})=\pm1.\notag
\eea
with $C_m=\pm1$. Therefore one choice to obtain a non-fractionalized topological paragmagnet is $C_{+1}=C_{-1}=-C_0=\pm1$. On the other hand, if all fermions share the same Chern number $C_{+1}=C_{-1}=C_0=\pm1$, there will be fractional charge polarization $\delta\rho_{-1}=\mp1/3$ and therefore 3-fold ground state degeneracy on torus.


\section{{Edge excitations protected by $U(1)_{S^z}$ symmetry}}

Since the effective Chern-Simons theory (\ref{lagrangian:S=1}) for the spin-1 state is obtained, its edge theory is straightforwardly obtained from bulk-edge correspondence\cite{Wen1995}. One can easily show\cite{Lu2012a} that no matter what perturbations are added to the system, there will be gapless edge excitations as long as $U(1)_{S^z}$ symmetry is preserved on the edge. The effective edge theory (the edge is along $\hat{x}$-direction) derived from bulk Chern-Simons theory (\ref{lagrangian:S=1}) is
\begin{eqnarray}
&\notag\mathcal{L}_{edge}=\frac{1}{4\pi}\sum_{I,J=\pm1}\big({\bf K}_{I,J}\partial_t\phi_I\partial_x\phi_J-{\bf V}_{I,J}\partial_x\phi_I\partial_x\phi_J\big)\\
&-\frac{1}{2\pi}\sum_{m\neq0}m\big(A_0^s\partial_x\phi_m-A_x^s\partial_t\phi_m\big).\notag
\end{eqnarray}
where ${\bf V}_{I,J}$ is a positive-definite real symmetric matrix which determines the velocity of edge modes. The $S^z$ density for the spin-1 system on the edge is given by
\begin{eqnarray}\notag
{S^z}(x)\simeq\sum_{m=\pm1}m\frac{\partial_x\phi_m(x)}{2\pi}.
\end{eqnarray}
and this $S^z$ component is in general gapped in the presence of $U(1)_{S^z}$ symmetry. On the other hand, the transverse components of spin-1 on the edge become
\begin{eqnarray}\notag
&S^+(x)=S^x(x)+\imth S^y(x)\simeq\\
&e^{\pm\imth(2\phi_{+1}+\phi_{-1})}+e^{\mp\imth(\phi_{+1}+2\phi_{-1})}.\notag
\end{eqnarray}
for ${\bf K}=\mp\begin{pmatrix}0&1\\1&0\end{pmatrix}$ in (\ref{K mat:S=1}). The commutation relation $[S^z(x),S^+(y)]=\delta(x-y)S^+(y)$ is implied by the Kac-Moody algebra $[\phi_{+1}(x),\partial_{x}\phi_{-1}(y)]=[\phi_{-1}(x),\partial_{x}\phi_{+1}(y)]=\mp2\pi\imth\delta(x-y)$ of chiral bosons $\{\phi_{\pm1}\}$ on the edge. The transverse components constitute the gapless excitations on the edge. Under a spin rotation along $\hat{z}$-axis the chiral bosons transform as
\begin{eqnarray}
&e^{\imth\theta\int S^z(y)\text{d}y}\phi_{m}(x)e^{-\imth\theta\int S^z(y)\text{d}y}=\phi_{m}(x)\pm m\theta.\notag
\end{eqnarray}
where $m=\pm1$ in the spin-1 system. If $U(1)_{S^z}$ symmetry is preserved, the edge excitations $\{\phi_{\pm1}(x)\}$ must be gapless and transverse spin components $\{S^\pm(x)\}$ have power-law correlations $\langle S^+(x,t)S^-(0,t)\rangle\sim1/|x|^4$. These edge states will remain gapless unless $U(1)_{S^z}$ symmetry is spontaneously broken and transverse magnetic order is developed on the edge\cite{Lu2012a}. On the other hand, the correlation functions of $S^z$ component also decay in a power-law fashion but with a different exponent: $\langle S^z(x,t)S^z(0,t)\rangle\sim1/|x|^2$.


\begin{figure}
 \includegraphics[width=0.3\textwidth]{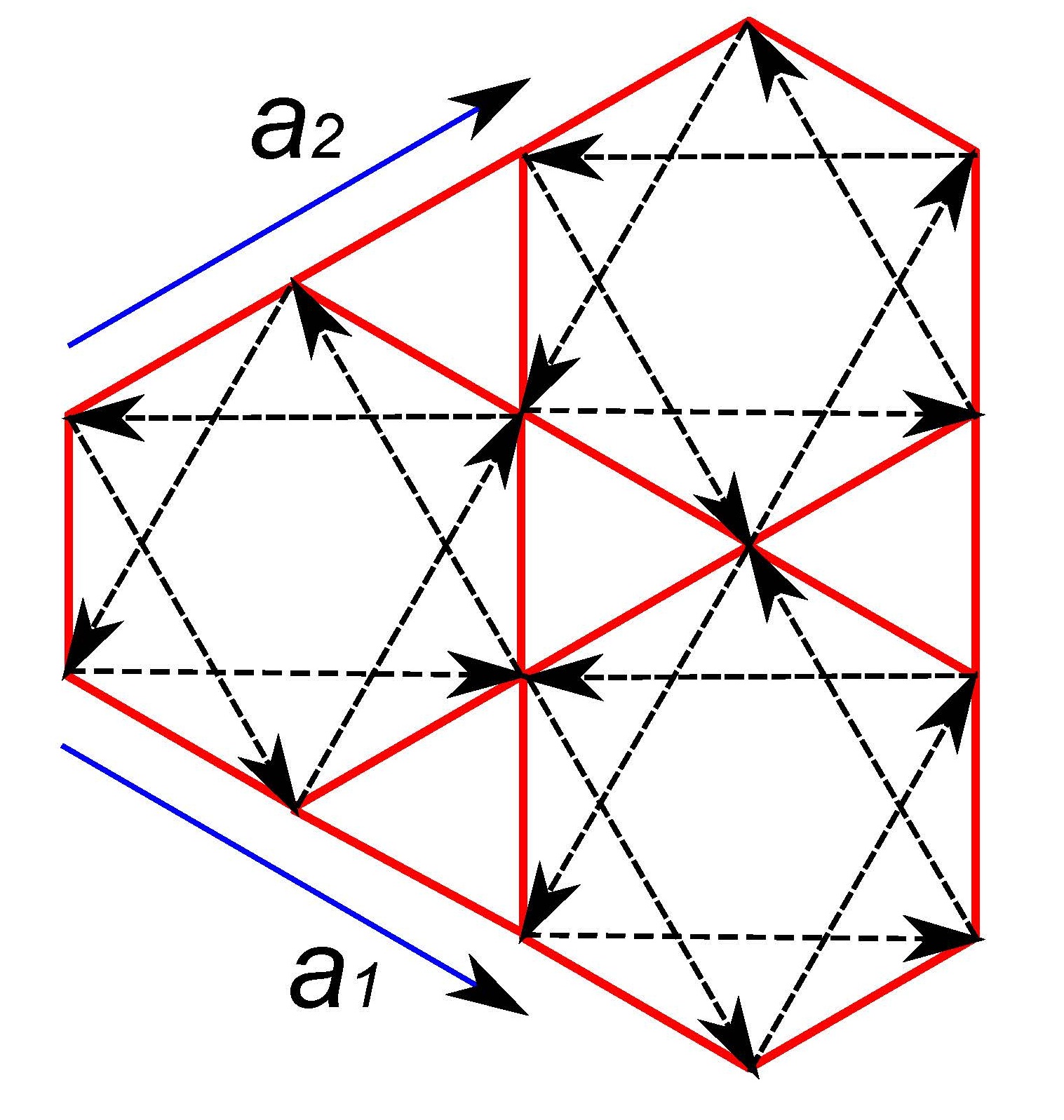}
\caption{(color online) Hopping Hamiltonian (\ref{kagome:mean-field Hamiltonian}) for $f_m$-fermion of spin-1 magnets on kagome lattice. Solid red lines are 1st nearest neighbor (NN) hopping terms with \emph{real} amplitude $t_1^m$, while dashed arrows represent 2nd NN hoppings with \emph{imaginary} amplitude $\imth t_2^m\cdot C_m\nu_{{\bf r}{\bf r}^\prime}$ between site ${\bf r}$ and ${\bf r}^\prime$. Here $\nu_{{\bf r}{\bf r}^\prime}=+1$ along on the arrow direction, or $-1$ along the opposite direction. $C_m=\pm1$ is the Chern number of the lowest band of this hopping Hamiltonian. $\vec{a}_{1,2}$ denote the two Bravais lattice vectors of kagome lattice and there are three lattice sites in each unit cell.}\label{fig:kagome hopping}
\end{figure}


\section{{Examples on kagome lattice and microscopic Hamiltonians}}

Here we demonstrate the above general construction by an explicit example: $U(1)_{S^z}$-symmetric spin-1 states on kagome lattice with spin quantum Hall effects. Microscopic Hamiltonians which could realize them are also derived from strong-coupling expansion\cite{Motrunich2007,McGreevy2012}.

\begin{figure}
 \includegraphics[width=0.5\textwidth]{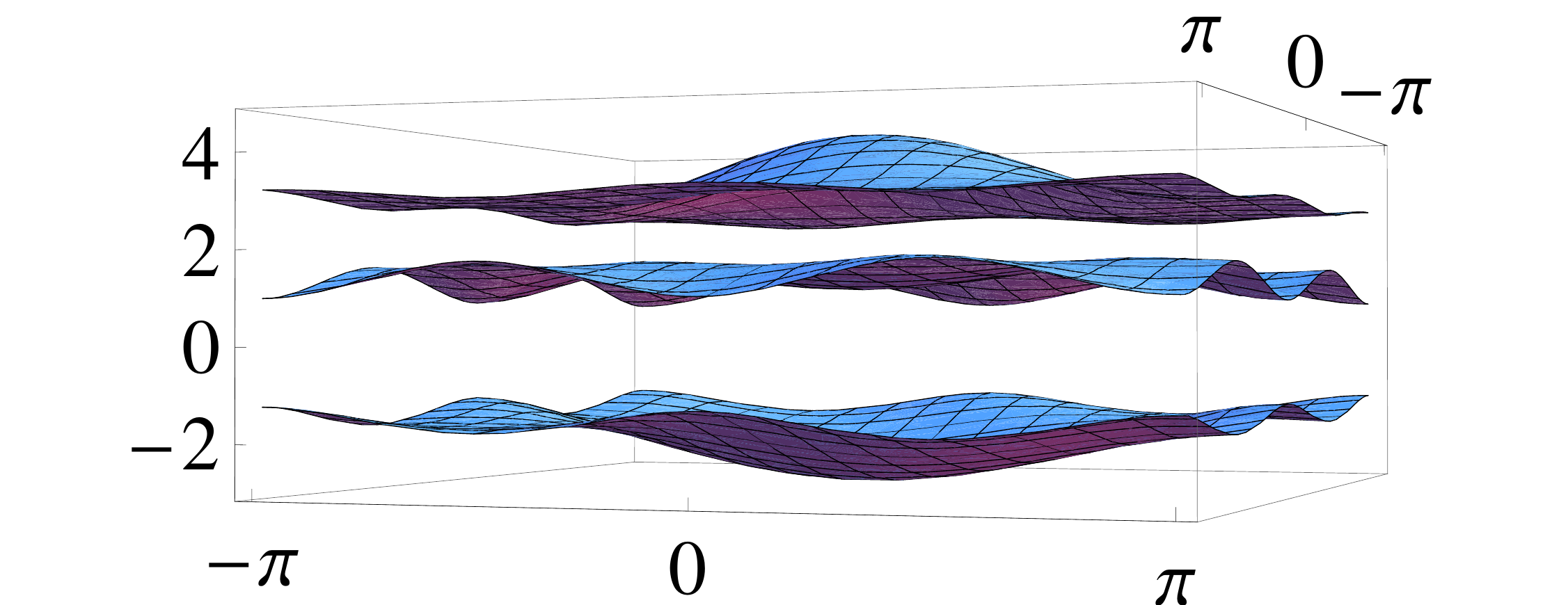}
\caption{(color online) Band structure of $f_m$-fermion hopping Hamiltonian (\ref{kagome:mean-field Hamiltonian}) in the 1st Brillouin zone of kagome lattice. The hopping parameters are chosen as $\mu^m=t_1^m=-1$ and $t_2^m=-1/2$. The Chern number for the three bands are $\{C_m,-2C_m,C_m\}$ from bottom to top, with the choice $C_m=\pm1$.}\label{fig:kagome band}
\end{figure}

The fermion hoping Hamiltonian (\ref{mean-field Hamiltonian}) for the $U(1)_{S^z}$-symmetric spin-1 states is shown in FIG. \ref{fig:kagome hopping}:
\begin{eqnarray}
&\label{kagome:mean-field Hamiltonian} H_{f}=\sum_{m=-1}^{+1}\Big(t_1^m\sum_{\langle{\bf r}{\bf r}^\prime\rangle}f^\dagger_{{\bf r},m}f_{{\bf r}^\prime,m}\\
&-\mu^m\sum_{{\bf r}}f^\dagger_{{\bf r},m}f_{{\bf r},m}+\imth t_2^m\cdot C_m\sum_{\langle\langle{\bf r}{\bf r}^\prime\rangle\rangle}\nu_{{\bf r}{\bf r}^\prime}f^\dagger_{{\bf r},m}f_{{\bf r}^\prime,m}\Big).\notag
\end{eqnarray}
where $\langle{\bf r}{\bf r}^\prime\rangle$ and $\langle\langle{\bf r}{\bf r}^\prime\rangle\rangle$ represent 1st and 2nd nearest neighbor pairs. Here $t_1^m,t_2^m<0$ are real hopping parameters and $\mu^m$ is the chemical potential for $f_m$-fermions. $C_m=\pm1$ is the Chern number of the lowest $f_m$-fermion band (see FIG. \ref{fig:kagome band}), which changes sign under time reversal. In the fermion state $|fermion\rangle$ in (\ref{projection:S})-(\ref{wf:S}) for spin-1 magnets, we require each species of fermions $\{f_m|m=0,\pm1\}$ to fill their lowest band of hopping Hamiltonian (\ref{kagome:mean-field Hamiltonian}). This can be done by choosing the chemical potential $\mu^m=t_1^m<0$ in (\ref{kagome:mean-field Hamiltonian}). One can see that on average there are 3 fermions per unit cell, or one fermion per site in $|fermion\rangle$, consistent with constraint (\ref{constraint:S=1}) for fermion representation (\ref{fermion rep:S=1}) of spin-1. When we choose the Chern number of the filled lowest bands to be $C_{+1}=C_{-1}=-C_0=\pm1$, after projection (\ref{projection:S})-(\ref{wf:S}) we obtain a non-fractionalized featureless spin-1 state with spin quantum Hall conductance $\sigma_{xy}^s=\pm2$. Such a featureless state preserves all the lattice symmetries of kagome lattice\footnote{Here the usual mirror reflection symmetry is defined in a slightly different way: a mirror reflection operation along \eg the horizontal line in FIG. \ref{fig:kagome hopping} is followed by an anti-unitary time reversal operation.} as well as $U(1)_{S^z}$ spin rotational symmetry along the $\hat{z}$-direction. However it does break time reversal symmetry, by having a quantized spin Hall conductance. It has no magnetization along $S^z$ direction, though.

The above construction on kagome lattice can be generalized to other two-dimensional lattices, as long as filled fermion bands have Chern numbers $C_{+1}=C_{-1}=-C_0=\pm1$ and one fermion per site on average. On square lattice \eg, where there is one site per unit cell, in general we need to insert flux in each plaquette\cite{Lu2012,McGreevy2012} for fermion hopping Hamiltonian (\ref{mean-field Hamiltonian}) to satisfy the requirement for Chern numbers. Although flux insertion breaks the explicit lattice translation symmetry for fermion hopping Hamiltonian, the spin-1 wavefunction (\ref{projection:S})-(\ref{wf:S}) after projection still preserves lattice symmetries.

The featureless spin-1 state $|spin\rangle$ is obtained from projection (\ref{projection:S})-(\ref{wf:S}) on the fermion state $|fermion\rangle$. This ``hard" projection can be implemented in a softer way, by introducing an energy penalty term for violating on-site constraint (\ref{constraint:S=1}) into fermion Hamiltonian (\ref{mean-field Hamiltonian}). Specifically the following four-fermion interaction term
\begin{eqnarray}
H_{int}=U\sum_{\bf r}\Big(\sum_{m=-1}^{+1}f^\dagger_{{\bf r},m}f_{{\bf r},m}-1\Big)^2=U\sum_{\bf r}\Big(\hat{N}_{{\bf r},f}-1\Big)^2.\notag
\end{eqnarray}
can enforce the constraint $\hat{N}_{{\bf r},f}=1$ in its low-energy subspace. In the strong-coupling limit $|t_{1,2}^m|\ll U$, the effective Hamiltonian in the zero-energy subspace of $H_{int}$ can be derived from degenerate perturbation theory:
\begin{eqnarray}
&\notag H_{eff}={\bf P}H_{f}\Big(1+{\bf Q}\frac{1}{H_{int}}{\bf Q}H_f\Big)^{-1}{\bf P}\\
&={\bf P}H_{f}{\bf P}-{\bf P}H_{f}{\bf Q}\frac{1}{H_{int}}{\bf Q}H_f{\bf P}+O(\frac{t^3}{U^2})\notag
\end{eqnarray}
where ${\bf P}$ is the projection operator onto the low-energy subspace fulfilling the single-occupancy constraint per site. ${\bf Q}=1-{\bf P}$ projects into the high-energy subspace where constraint (\ref{constraint:S=1}) is violate. The low-energy subspace satisfying single-occupancy constraint is nothing but the spin-1 Hilbert space and we obtain a spin-1 Hamiltonian which favors the \emph{projected} groundstate of $H_f$ as its groundstate.

For the specific fermion model (\ref{kagome:mean-field Hamiltonian}) on kagome lattice, the corresponding effective spin-1 model derived from strong-coupling perturbation theory has the following form (up to order $\sim t^4/U^3$ terms):
\begin{eqnarray}
\notag H_{kagome}=\sum_{\bf r}\Tr{\big(\Lambda-4\frac{{\bf T}_1{\bf T}_1^\ast+{\bf T}_2{\bf T}_2^\ast}{U}\big)\hat{\bf M}_{\bf r}}\\
\notag+\frac{1}{U}\sum_{\langle{\bf r}{\bf r}^\prime\rangle}\Tr{{\bf T}_1\hat{\bf M}_{\bf r}{\bf T}_1^\ast\hat{\bf M}_{{\bf r}^\prime}}+\frac{1}{U}\sum_{\langle\langle{\bf r},{\bf r}^\prime\rangle\rangle}\Tr{{\bf T}_2\hat{\bf M}_{\bf r}{\bf T}_2^\ast\hat{\bf M}_{{\bf r}^\prime}}\\
\notag+\frac1{U^2}\Big(\sum_{\triangpic}\mathcal{R}_{{\bf r}_1,{\bf r}_2,{\bf r}_3}+\sum_{\nnring}\mathcal{R}_{{\bf r}_1,{\bf r}_2,{\bf r}_3}+\sum_{\nnnring}\mathcal{R}_{{\bf r}_1,{\bf r}_2,{\bf r}_3}\Big)\\
\label{spin-1 model:kagome}
\end{eqnarray}
where $\Lambda=\text{diag}(\mu^m)\equiv\text{diag}(\mu^{+1},\mu^0,\mu^{-1})$ is the chemical potential matrix, while ${\bf T}_1=\text{diag}(t_1^m)$ and ${\bf T}_2=\text{diag}(\imth C_mt_2^m)$ are 1st and 2nd NN fermion hopping matrices. Matrix $\hat{\bf M}_{\bf r}$ is defined in terms of local spin operators (or on-site fermion bilinears):
\begin{eqnarray}
\Big(\hat{\bf M}_{\bf r}\Big)_{l,m}\equiv f_{{\bf r},l}f^\dagger_{{\bf r},m}=\begin{pmatrix}\frac{2-S^z_{\bf r}(S^z_{\bf r}+1)}2&\frac{-S^-_{\bf r}S^z_{\bf r}}{\sqrt2}
&\frac{-(S^-_{\bf r})^2}2\\ \frac{-S^z_{\bf r}S^+_{\bf r}}{\sqrt2}&(S^z_{\bf r})^2&\frac{S^z_{\bf r}S^-_{\bf r}}{\sqrt2} \\ \frac{-(S_{\bf r}^+)^2}2
&\frac{S^+_{\bf r}S^z_{\bf r}}{\sqrt2}&\frac{2-S^z_{\bf r}(S^z_{\bf r}-1)}2\end{pmatrix}.\notag
\end{eqnarray}

The 1st line of this Hamiltonian represents on-site $S^z$ anisotropy terms, compatible with the $U(1)_{S^z}$ spin rotational symmetry. The 2nd line are two- and four-spin interactions between 1st and 2nd NNs. They reduce to the $SU(3)$ bilinear-biquadratic form $\vec{S}_{\bf r}\cdot\vec{S}_{{\bf r}^\prime}+(\vec{S}_{\bf r}\cdot\vec{S}_{{\bf r}^\prime})^2$, when three species of fermions $\{f_0,f_{\pm1}\}$ have exactly the same hopping parameters in (\ref{kagome:mean-field Hamiltonian}). The 3rd line corresponds to ring exchange terms in three different types of 1st/2nd NN triangles:
\begin{eqnarray}
&\notag\mathcal{R}_{{\bf r}_1,{\bf r}_2,{\bf r}_3}=\Tr{\hat{\bf M}_{{\bf r}_2}{\bf T}_{{\bf r}_2{\bf r}_3}\hat{\bf M}_{{\bf r}_3}{\bf T}_{{\bf r}_3{\bf r}_1}\big({\bf E}-\hat{\bf M}_{{\bf r}_1}\big){\bf T}_{{\bf r}_1{\bf r}_2}}.
\end{eqnarray}
where ${\bf E}$ represents a $3\times3$ identity matrix. ${\bf T}_{{\bf r}{\bf r}^\prime}$ denotes the $3\times3$ fermion hopping matrix from site ${\bf r}$ to ${\bf r}^\prime$. Under time reversal operation, the spin-1 operators transform as
\begin{eqnarray}
&\hat{\bf M}_{\bf r}\rightarrow\begin{pmatrix}0&0&1\\0&-1&0\\1&0&0\end{pmatrix}\hat{\bf M}_{\bf r}\begin{pmatrix}0&0&1\\0&-1&0\\1&0&0\end{pmatrix}.\notag
\end{eqnarray}
It's straightforward to show that imaginary 2nd NN hoppings with Chern number $C_{+1}=C_{-1}=-C_0=\pm1$ and $t_2^m<0$ will lead to no time reversal symmetry in the spin-1 Hamiltonian $H_{kagome}$. Both the 2nd and 3rd line in model (\ref{spin-1 model:kagome}) break the time reversal symmetry. On the other hand, this spin-1 Hamiltonian does have $U(1)_{S^z}$ spin rotational symmetry, and all the kagome lattice symmetries.

From strong-coupling perturbation theory, we only derive the spin-1 Hamiltonian $H_{kagome}$ up to $O(t^4/U^3)$ terms in $t/U$ expansion. A careful study of spin-1 Hamiltonian (\ref{spin-1 model:kagome}) and the effects of higher-order terms in perturbation expansion are beyond the scope of this paper and we leave them for future studies. 



\section{Conclusions}

In this work we show that a class of gapped featureless non-fractionalized spin-1 states, \ie spin-1 topological paramagnets, in analogy to AKLT state in a spin-1 chain, can be constructed on a generic two-dimensional lattice. Using the fermion representation of integer-spins we can write down the many-body wavefunction of these states, and derive microscopic Hamiltonians which may realize them on the kagome lattice. The bulk effective theory of these gapped phases are obtained, along with gapless edge excitations protected by $U(1)_{S^z}$ symmetry and spin quantum Hall conductance $\sigma_{xy}=\pm2$ in unit of $\hbar/2\pi$. A concrete example is shown for spin-1 magnets on the two-dimensional kagome lattice, which preserves all the kagome lattice symmetries and has no magnetization.\\

\acknowledgements

YML thanks Ashvin Vishwanath for helpful discussions, and Kavli Institute for Theoretical Physics for hospitality, where part of this work was finished during 2012 program ``Frustrated Magnetism and Quantum Spin Liquids''. We acknowledge the support by the DOE grant number DE-AC02-05CH11231 (YML,DHL) and in part by the National Science Foundation under Grant No. NSF PHY11-25915(YML).


\end{document}